\documentstyle[epsfig,12pt,a4p]{article}

\newcommand{\etaeta}{\mbox{$\eta \eta$ }}
\newcommand{\etaetap}{\mbox{$\eta \eta^\prime$ }}
\newcommand{\etapetap}{\mbox{$\eta^\prime \eta^\prime$ }}

\begin{document}
\begin{titlepage}
\def\footnoterule{\hrule width 1.0\columnwidth}
\begin{tabbing}
put this on the right hand corner using tabbing so it looks
 and neat and in \= \kill
\> {28 March 2000}
\end{tabbing}
\bigskip
\bigskip
\begin{center}{\Large  {\bf A study of the
\etaeta channel produced in central
pp interactions at 450 GeV/c}
}\end{center}

\bigskip
\bigskip
\begin{center}{        The WA102 Collaboration
}\end{center}\bigskip
\begin{center}{
D.\thinspace Barberis$^{  4}$,
F.G.\thinspace Binon$^{   6}$,
F.E.\thinspace Close$^{  3,4}$,
K.M.\thinspace Danielsen$^{ 11}$,
S.V.\thinspace Donskov$^{  5}$,
B.C.\thinspace Earl$^{  3}$,
D.\thinspace Evans$^{  3}$,
B.R.\thinspace French$^{  4}$,
T.\thinspace Hino$^{ 12}$,
S.\thinspace Inaba$^{   8}$,
A.\thinspace Jacholkowski$^{   4}$,
T.\thinspace Jacobsen$^{  11}$,
G.V.\thinspace Khaustov$^{  5}$,
J.B.\thinspace Kinson$^{   3}$,
A.\thinspace Kirk$^{   3}$,
A.A.\thinspace Kondashov$^{  5}$,
A.A.\thinspace Lednev$^{  5}$,
V.\thinspace Lenti$^{  4}$,
I.\thinspace Minashvili$^{   7}$,
J.P.\thinspace Peigneux$^{  1}$,
V.\thinspace Romanovsky$^{   7}$,
N.\thinspace Russakovich$^{   7}$,
A.\thinspace Semenov$^{   7}$,
P.M.\thinspace Shagin$^{  5}$,
H.\thinspace Shimizu$^{ 10}$,
A.V.\thinspace Singovsky$^{ 1,5}$,
A.\thinspace Sobol$^{   5}$,
M.\thinspace Stassinaki$^{   2}$,
J.P.\thinspace Stroot$^{  6}$,
K.\thinspace Takamatsu$^{ 9}$,
T.\thinspace Tsuru$^{   8}$,
O.\thinspace Villalobos Baillie$^{   3}$,
M.F.\thinspace Votruba$^{   3}$,
Y.\thinspace Yasu$^{   8}$.
}\end{center}

\begin{center}{\bf {{\bf Abstract}}}\end{center}

{
The reaction
$ pp \rightarrow p_{f} (\eta \eta ) p_{s}$ has been studied at 450 GeV/c.
For the first time a
partial wave analysis of the centrally produced \etaeta system has been
performed. Signals for the $f_0(1500)$, $f_0(1710)$ and $f_2(2150)$
are observed and the decay branching fractions of these states are determined.
}
\bigskip
\bigskip
\bigskip
\bigskip\begin{center}{{Submitted to Physics Letters}}
\end{center}
\bigskip
\bigskip
\begin{tabbing}
aba \=   \kill
$^1$ \> \small
LAPP-IN2P3, Annecy, France. \\
$^2$ \> \small
Athens University, Physics Department, Athens, Greece. \\
$^3$ \> \small
School of Physics and Astronomy, University of Birmingham, Birmingham, U.K. \\
$^4$ \> \small
CERN - European Organization for Nuclear Research, Geneva, Switzerland. \\
$^5$ \> \small
IHEP, Protvino, Russia. \\
$^6$ \> \small
IISN, Belgium. \\
$^7$ \> \small
JINR, Dubna, Russia. \\
$^8$ \> \small
High Energy Accelerator Research Organization (KEK), Tsukuba, Ibaraki 305-0801,
Japan. \\
$^{9}$ \> \small
Faculty of Engineering, Miyazaki University, Miyazaki 889-2192, Japan. \\
$^{10}$ \> \small
RCNP, Osaka University, Ibaraki, Osaka 567-0047, Japan. \\
$^{11}$ \> \small
Oslo University, Oslo, Norway. \\
$^{12}$ \> \small
Faculty of Science, Tohoku University, Aoba-ku, Sendai 980-8577, Japan. \\
\end{tabbing}
\end{titlepage}
\setcounter{page}{2}
\bigskip
\par
The \etaeta, \etaetap and \etapetap channels are important
for the determination of the gluonic content of mesons, since
it is thought likely that glueballs will decay
with the emission of
$\eta$s and $\eta^{\prime}$s~\cite{re:gers,re:b}.
In addition,
central production is proposed as a good place to produce
glueballs via Double Pomeron Exchange (DPE)~\cite{re:b,re:bb}.
Hence the interest of the present paper which studies
the centrally produced \etaeta system.
\par
The \etaeta channel has been studied previously
in radiative $J/\psi$ decays
by the Crystal Ball experiment~\cite{cball},
in $\pi^-p$ interactions by the NA12 experiment~\cite{NA12h,konda},
in central production by NA12/2~\cite{na12etaeta} and in
$p \overline p$ annihilations by the Crystal Barrel
Collaboration~\cite{cb1} and E760 at Fermilab~\cite{E760}.
In all these reactions evidence for the $f_0(1500)$ has emerged.
In addition, in $p \overline p$ annihilation
evidence is claimed
for a $f_0(2100)$~\cite{E760,cb2}
and an \etaeta decay mode of the $f_2(1950)$~\cite{cb2}.
In $\pi^-p$ interactions the NA12 experiment claim a $f_0(2000)$~\cite{konda}.
In radiative $J/\psi$ decays evidence was reported for the
$f_J(1710)$~\cite{cball}.
In central production the NA12/2 experiment claimed
evidence for a $f_2(2175)$~\cite{na12etaeta}.
\par
In this paper a study is presented of the \etaeta
final state formed in the reaction
\begin{equation}
pp \rightarrow p_{f} (\eta \eta) p_{s}
\label{eq:e}
\end{equation}
at 450~GeV/c.
It represents more than a factor of 6 increase in statistics over the only
other data on the centrally produced \etaeta final state~\cite{na12etaeta}
and moreover will present the first partial wave analysis of this
channel in central production.
The data come from the WA102 experiment
which has been performed using the CERN Omega Spectrometer,
the layout of which is
described in ref.~\cite{WADPT}.
Reaction~(\ref{eq:e})
has been isolated using the following decay modes:
\[
\begin{array}{cc}
\eta \rightarrow \gamma \gamma &\eta \rightarrow \gamma \gamma \\
\eta \rightarrow \gamma \gamma &\eta \rightarrow \pi^+\pi^-\pi^0 \\
\eta \rightarrow \pi^+\pi^-\pi^0 &\eta \rightarrow \pi^+\pi^-\pi^0 \\
\end{array}
\]
The above decay modes account for 38.8~\% of the total.
\par
Fig.~\ref{fi:1}a) shows a scatter plot of $M(\gamma \gamma)$ versus
$M(\gamma \gamma)$
which has been extracted from the sample of events having
two outgoing
charged tracks and four $\gamma$s reconstructed in the GAMS-4000
calorimeter using momentum and energy balance.
A clear signal of the \etaeta channel can be observed.
Fig.~\ref{fi:1}b) shows the $\gamma \gamma$ mass spectrum if
the other $\gamma \gamma$ pair is compatible with being an $\eta$
(0.48~$\leq$~M($\gamma \gamma$)~$\leq$~0.62~GeV) where a clear $\eta$ signal
can
be observed.
The \etaeta final state has been selected using the mass cuts described above.
\par
Fig.~\ref{fi:1}c) shows a scatter plot of $M(\gamma \gamma)$ versus
M($\pi^+\pi^-\pi^0)$ for the
sample of events having four
outgoing
charged tracks and four $\gamma$s reconstructed in the GAMS-4000
calorimeter after imposing momentum and energy balance.
A clear signal of the \etaeta channel can be observed.
Fig.~\ref{fi:1}d) shows the $\pi^+\pi^-\pi^0$ mass spectrum if
the $\gamma \gamma$ mass is compatible with being an $\eta$
(0.48~$\leq$~M($\gamma \gamma$)~$\leq$~0.62~GeV)
and fig~\ref{fi:1}e) shows the $\gamma \gamma$ mass spectrum if the
$\pi^+\pi^-\pi^0$ mass is compatible with being an $\eta$
(0.52~$\leq$~M($\pi^+\pi^-\pi^0$)~$\leq$~0.58~GeV).
The \etaeta final state has been selected by requiring that
0.48~$\leq$~M($\gamma \gamma$)~$\leq$~0.62~GeV and
0.52~$\leq$~M($\pi^+\pi^-\pi^0$)~$\leq$~0.58~GeV.
\par
Fig.~\ref{fi:1}f) shows a scatter plot of $M(\pi^+\pi^-\pi^0)$ versus
$M( \pi^+\pi^-\pi^0)$ for the
the sample of events having six
outgoing
charged tracks and four $\gamma$s reconstructed in the GAMS-4000
calorimeter after imposing momentum and energy balance.
A signal of the \etaeta channel can be observed.
Fig.~\ref{fi:1}g) shows the $\pi^+\pi^-\pi^0$ mass spectrum if
the other $\pi^+\pi^-\pi^0$ combination has a mass compatible with being an
$\eta$
(0.52~$\leq$~M($\pi^+\pi^-\pi^0$)~$\leq$~0.58~GeV).
The \etaeta final state has been selected using the mass cuts described above.
\par
The background below the $\eta$ signal has several sources
including combinatorics, fake gammas and other channels.
The combinatorial background is removed, in part,
in the selection procedure. The
remaining background varies from 16 \% to 32 \% dependent on the
decay topology.
Three methods have been used to determine the effects of this background;
studying the side bands around the $\eta$ signal, studying events that
do not balance momentum and using event mixing.
All three methods give a similar background which peaks near threshold
in the \etaeta mass spectrum. To illustrate this point, superimposed
on figs.~\ref{fi:1}b), d), e) and g) as a shaded histogram are
the respective mass distributions when the central system has a
mass greater than 1.3~GeV. As can be seen the background below the
$\eta$ signal is reduced with respect to the total sample.
In the remainder of this paper the method used to determine
the background will be the one using events that do not balance momentum.
The resulting distributions well represent the
background in the data.
\par
The \etaeta mass spectra from each decay mode are
very similar and the combined mass spectrum is shown in
fig.~\ref{fi:2}a) and consists of 3351 events.
Superimposed on the mass spectrum as a shaded histogram is the
estimate of the background.
Fig.~\ref{fi:2}b) shows the background subtracted mass spectrum.
The mass spectrum has a threshold enhancement and has peaks at
1.5 and 2.1 GeV and a shoulder in the 1.7~GeV region.
Superimposed on fig~\ref{fi:2}b) as a shaded histogram is the expected
contribution
from the $f_2(1270)$ (47 events) and $f_2^\prime(1525)$ (3 events)
using the observed $f_2(1270)$ signal in the $\pi^+\pi^-$ final
state~\cite{pipipap}, the observed
$f_2^\prime(1525)$ signal in the $K^+K^-$ final state~\cite{kkpap}
and
correcting
for the experimental acceptance and using
the
PDG~\cite{PDG98}
branching ratios.
As can be seen the majority of the signal at 1.5~GeV is not due to the
$f_2^\prime(1525)$.
\par
A Partial Wave Analysis (PWA) of the centrally produced \etaeta system
has been
performed assuming the \etaeta system is produced by the
collision of two particles (referred to as exchanged particles) emitted
by the scattered protons.
The $z$ axis
is defined by the momentum vector of the
exchanged particle with the greatest four-momentum transferred
in the \etaeta centre of mass.
The $y$ axis is defined
by the cross product of the momentum vectors of
the two exchanged particles in the $pp$ centre of mass.
The two variables needed to specify the decay process were taken as the polar
and azimuthal angles ($\theta$, $\phi$) of one
of the $\eta$s in the \etaeta
centre of mass relative to the coordinate system described above.
\par
The acceptance corrected moments $\sqrt{4\pi}t_{LM}$, defined by
\[
I(\Omega) =  \sum_L t_{L0} Y^0_L(\Omega) +
2 \sum_{L,M >0} t_{LM}Re\{Y^M_L(\Omega)\}
\]
have been rescaled to the total number of observed events and
are shown in fig.~\ref{fi:3}. The moments
with \hbox{$M>2$}
and all the moments
with \hbox{$L>4$}
(e.g. the $t_{44}$ and $t_{60}$ moments shown in fig.~\ref{fi:3})
are small and hence only
partial waves with spin \hbox{$l=0$} and 2 and absolute values of spin
$z$-projection \hbox{$m\leq1$} have been included in the PWA.
\par
The amplitudes used for the PWA are defined in the reflectivity
basis~\cite{reflectivity}.
In this basis the angular distribution is given by a sum of two non-interfering
terms corresponding to negative and positive values of reflectivity.
The waves used were of the form $J^\varepsilon _m$ with $J$~$=$~$S$
and $D$,
$m$~$=$~$0,1$ and reflectivity $\varepsilon$~=~$\pm 1$.
The expressions relating the moments
($t_{LM}$) and the waves ($J^\varepsilon _m$) are given in
ref.\cite{pi0pi0pap}.
Since the overall phase for each reflectivity is indeterminate,
one wave in each reflectivity can be set to be real ($S_0^-$ and $D_1^+$
for example) and hence two phases can be set to zero ($\phi_{S_0^-}$ and
$\phi_{D_1^+}$ have been chosen).
This results in 6 parameters to be determined from the fit to the
angular distributions.
\par
The PWA has been performed independently in 80~MeV intervals of the \etaeta
mass spectrum. In each mass an event-by-event maximum likelihood
method has been used. The function
\begin{equation}
F=-\sum_{i=1}^Nln\{I(\Omega)\} + \sum_{L,M}t_{LM}\epsilon_{LM}
\label{eq:2}
\end{equation}
has been minimised, where $N$ is the number of events in a given mass bin,
$\epsilon_{LM}$ are the efficiency corrections calculated
in the centre of the bin
and $t_{LM}$ are the moments of the angular distribution.
The moments calculated from the partial amplitudes
are shown superimposed as histograms on the experimental moments
in fig~\ref{fi:3}. As can be seen the results of the fit
reproduce well the experimental moments.
\par
The system of equations which express the moments via the partial wave
amplitudes is non-linear which leads to inherent
ambiguities. For a system with S and D waves there are two solutions
for each mass bin.
In each mass bin
one of these solutions
is found from the fit to the experimental angular distributions,
the other one can then be calculated by the method described in
ref.~\cite{reflectivity}.
In the case under study the bootstrapping procedure is trivial
because the Barrelet
function has only two roots ($Z_i$ with $i$~=~1,2) and their real and
imaginary parts do not cross zero as functions of mass,
as seen in fig~\ref{fi:2}c) and d).
In order to link the solutions in adjacent mass bins,
the real parts of the roots are
sorted in each mass bin in such a way that
the real part of the first root should be
larger than the real part of the second root
(real parts of the two roots have different signs).
For the first solution the imaginary parts of both roots are
taken positive, the second solution is obtained by
complex conjugation of one of the roots.
\par
In this case two different
PWA solutions are found. One solution is dominated by the S-wave
the other solution has the events split between the different
D-waves.
By definition both solutions give identical moments and identical
values of the likelihood.
A Monte Carlo study has shown that if the input distribution
is really dominated
by a D-wave (i.e. $D^-_0$, $D^-_1$ or $D^+_1$)
then both solutions of the PWA will be dominated by
that D-wave.
However, if the input distribution is dominated by the S-wave then
two different solutions will result.
One of the PWA solutions will be dominated by the
S-wave, the other solution
will be split equally between
the D-waves and hence the physical solution
can still be determined.
The physical solution is shown in fig.~\ref{fi:4}.
The $S_0^-$-wave for the
physical solution is characterised by a broad enhancement
at threshold and
a peak at 1.5~GeV. A broad enhancement is
also seen in the $D^-_0$-wave at 2.1~GeV.
Superimposed on the waves as a histogram
is the result of running the
PWA on the background events.
\par
Fig.~\ref{fi:5} shows the background subtracted $S^-_0$ and $D^-_0$ waves.
The PWA analysis has also been performed by extending the
likelihood function given in equation~(\ref{eq:2}) to include the
background subtraction, namely
\[
F=-\sum_{i=1}^Nln\{I(\Omega)\} + \sum_{L,M}t_{LM}\epsilon_{LM}
+\sum_{i=1}^{N_{bg}}ln\{I(\Omega)\}
\]
where $N_{bg}$ is the number of background events. Since the
background is dominantly S-wave the results of this method
are similar to those shown in fig.~\ref{fi:5}.
\par
The $S^-_0$-wave has been fitted using a K-matrix
parameterisation similar to that
used to fit the $K^+K^-$ spectrum~\cite{pipikkpap} with the addition
of an incoherent background term.
Poles have been introduced to describe the $f_0(1370)$, $f_0(1500)$
and $f_0(1710)$ with parameters fixed to those found from the
coupled channel fit to the $\pi^+ \pi^-$ and
$K^+K^-$ spectra~\cite{pipikkpap}, namely
M($f_0(1370)$)~=~1312~MeV, $\Gamma(f_0(1370))$~=~218~MeV,
M($f_0(1500)$)~=~1502~MeV, $\Gamma(f_0(1500))$ = 98~MeV and
M($f_0(1710)$)~=~1727~MeV, $\Gamma(f_0(1710))$ = 126~MeV.
The fit describes well the data but, due to the size of the errors,
it is not possible to conclude about the need for
a further resonance above 2~GeV.
The inclusion of the $f_0(1710)$ is essential to describe the
spectrum.
\par
The $D_0^-$ wave has been fitted using two spin 2 relativistic Breit-Wigners
and a linear background. The first Breit-Wigner is used to describe the
$f_2(1270)$ with mass and width fixed to the PDG
values~\cite{PDG98} and the second to describe the peak at 2.1~GeV
where we have used M~=~2130~MeV, $\Gamma$~=~270~MeV which are
the parameters found for $f_2(2150)$ observed in the
$K^+K^-$ final state~\cite{kkpap}.
As can be seen the fit well describes the data.
This state is compatible with the $f_2(2175)$
previously observed by the NA12/2 experiment
in the \etaeta final state~\cite{na12etaeta}.
\par
The error bars introduced by the partial wave analysis
do not allow
the parameters of the resonances
to be determined
from a free fit to the
waves. Instead, we have performed a fit to the total
mass spectrum shown in fig.~\ref{fi:2}b) using an incoherent sum
of the expressions used to fit the $S^-_0$ and $D^-_0$ waves.
The parameters for the $f_0(1370)$ and $f_2(1270)$ have been fixed to the
values
used above.
The fit is shown superimposed on fig.~\ref{fi:2}b) and for the
scalar resonances yields
sheet II T-Matrix poles~\cite{sheet} at
\begin{tabbing}
0000aaaa\=adfsfsf99baxxx \=Ma \= = \=12249 \=pi \=1200 \=i22\=2400 \=pi
\=1200000  \=MeV   \kill
\>$f_0(1500)$ \>M
\>=\>(1510\>$\pm$\>$\;\;8$)\>$-i$\>($\;\;55$\>$\pm$\>$\;\;8$)\>MeV\\
\>$f_0(1710)$ \>M
\>=\>(1698\>$\pm$\>18)\>$-i$\>($\;\;60$\>$\pm$\>$13$)\>MeV
\end{tabbing}
These
parameters are consistent with the PDG~\cite{PDG98} values for these
resonances and our previous fits~\cite{pipikkpap}.
For the $f_2(2150)$ the fit gives
\begin{tabbing}
0000aaaa\=adfsfsf99baxxx \=Ma \= = \=1224 \=pm \=120 \=MeVsfw, \=gaa \=
= \=1224 \=pm \=120  \=MeV   \kill
\>$f_2(2150)$ \>M \>=\>2151\>$\pm$\>16\>MeV,\>$\Gamma$\>=\>280\>$\pm$\>70\>MeV.
\end{tabbing}
The parameters found are compatible with those from the $K^+K^-$
channel~\cite{kkpap}.
\par
After correcting for the unseen decay modes
and using data from the previously observed
$\pi \pi$~\cite{pipipap,pi0pi0pap} final state
the branching ratio \etaeta/$\pi \pi$
has been calculated for the $f_0(1500)$ from the
fit to the $S_0^-$ wave and gives  :
\[
\frac{f_0(1500) \rightarrow \eta\eta}{f_0(1500) \rightarrow \pi \pi}
= 0.18 \pm 0.03
\]
This value agrees well with the value that can be derived from the
PDG~\cite{PDG98} of 0.23~$\pm$~0.10.
\par
For the $f_2(1270)$ the ratio is
\[
\frac{f_2(1270) \rightarrow \eta\eta}{f_2(1270) \rightarrow \pi \pi}
= (3 \pm 1)\times 10^{-3}
\]
which is compatible with the PDG~\cite{PDG98} value.
\par
For the $f_0(1500)$, $f_0(1710)$ and $f_2(2150)$
the branching ratio \etaeta/$K \overline K$
has been calculated from the WA102 data to be:
\[
\frac{f_0(1500) \rightarrow \eta \eta} {f_0(1500) \rightarrow K \overline K}
 = 0.54 \pm 0.12
\]
\[
\frac{f_0(1710) \rightarrow \eta \eta} {f_0(1710) \rightarrow K \overline K}
 = 0.48 \pm 0.15
\]
\[
\frac{f_2(2150) \rightarrow \eta \eta} {f_2(2150) \rightarrow K \overline K}
 = 0.78 \pm 0.14
\]
The branching ratios for the $f_0(1500)$, $f_0(1710)$ and $f_2(2150)$
differ from that of a known $s \overline s$ state, the
$f_2^\prime(1525)$, which has a ratio of 0.12~$\pm$~0.04~\cite{PDG98}
which is consistent with the SU(3) prediction for such a state~\cite{proko}.
No signal is seen for the
$f_2^\prime(1525)$ in the \etaeta final state of WA102 and
an upper limit for its decay to this channel has been calculated, which gives
\[
\frac{f_2^\prime(1525) \rightarrow \eta\eta}
{f_2^\prime(1525) \rightarrow K \overline K}
< 0.14 \;\; (90 \% \;\;CL)
\]
which is compatible with the PDG~\cite{PDG98} value given above.
\par
For the $f_0(1370)$ there is considerable uncertainty due to the
background subtraction which mainly affects the $S_0^-$ wave at
threshold. The strongest $f_0(1370)$ signal has been observed in
the $4\pi$ final state therefore the branching ratio \etaeta /$4\pi$
has been calculated to be
\[
\frac{f_0(1370) \rightarrow \eta\eta}{f_0(1370) \rightarrow 4 \pi }
= (4.7 \pm 2)\times 10^{-3}
\]
which is compatible with the Crystal Barrel measurement of
(1.7~$\pm$~0.9)$\times 10^{-3}$~\cite{thoma}.
\par
In the D-waves there is no evidence for the
$f_2(1950)$ which has been claimed to have been seen in
the \etaeta final states formed in $p \overline p$
annihilations~\cite{cb2}. Since the $f_2(1950)$ is clearly seen in the
$4 \pi$ final state of experiment WA102~\cite{pi4pap}, an upper
limit for its decay to \etaeta has been calculated and gives
\[
\frac{f_2(1950) \rightarrow \eta\eta}{f_2(1950) \rightarrow  4\pi}
< 5.0\times 10^{-3}\;\;  (90 \% \;\;CL)
\]
Hence this would imply that if the observation in the \etaeta
final state of $p \overline p$ annihilations is correct, then
a very large signal for the $f_2(1950)$ should be seen in the
$4\pi$ final state of the same experiment.
\par
The $f_0(1500)$ has previously been observed in the
$\pi \pi$~\cite{pipipap,pi0pi0pap}, $K \overline K$~\cite{kkpap},
$4 \pi$~\cite{pi4pap} and $\eta \eta^\prime$~\cite{etaetappap} final states
of the WA102 experiment.
The relative decay rates
for the $f_0(1500) $ are calculated to be:
\[
\pi \pi : K \overline K : \eta\eta : \eta\eta^\prime : 4\pi
= 1 \;:\; 0.33 \pm 0.07\;
:\; 0.18 \pm 0.03\; :\; 0.096 \pm 0.026 \;
:\; 1.36 \pm 0.15
\]
The $f_0(1370)$ is below $\eta \eta^\prime$ threshold.
The remaining
relative decay rates
for the $f_0(1370)$
are:
\[
\pi \pi : K \overline K :  \eta\eta : 4\pi
= 1 : 0.46 \pm 0.19 : 0.16 \pm 0.07 : 34.0 ^{+22}_{-9}
\]
No signal was observed for the $f_0(1710)$ in the $4\pi$~\cite{pi4pap}
or the $\eta \eta^\prime$ channels~\cite{etaetappap}.
Therefore,
an upper limit has been calculated for its decay to these
final states.
For the $f_0(1710) $
the relative decay rates
are:
\[
\pi \pi : K \overline K :  \eta\eta : \eta\eta^\prime : 4\pi
= 1 : 5.0 \pm 0.7 : 2.4 \pm 0.6
: <\;0.18\;(90\;\%\;\; CL) : <\;5.4\;(90\;\%\;\; CL)
\]
\par
In summary,
a partial wave analysis of the \etaeta channel has been performed
for the first time in central production.
Clear evidence is found for an \etaeta decay mode of the $f_0(1500)$,
$f_0(1710)$ and $f_2(2150)$ and the decay branching ratios of these states
have been determined.
\begin{center}
{\bf Acknowledgements}
\end{center}
\par
This work is supported, in part, by grants from
the British Particle Physics and Astronomy Research Council,
the British Royal Society,
the Ministry of Education, Science, Sports and Culture of Japan
(grants no. 07044098 and 1004100), the French Programme International
de Cooperation Scientifique (grant no. 576)
and
the Russian Foundation for Basic Research
(grants 96-15-96633 and 98-02-22032).

\clearpage
{ \large \bf Figures \rm}
\begin{figure}[h]
\caption{
Selection of the \etaeta final state.
For the reaction $ pp \rightarrow p_fp_s(4 \gamma)$
a) M($\gamma \gamma$) versus M($\gamma \gamma$) and
b) M($\gamma \gamma$) if the other $\gamma \gamma$
pair is in the $\eta$ band
(0.48~$\leq$~M($\gamma \gamma$)~$\leq$~0.62~GeV).
For the reaction $ pp \rightarrow p_fp_s(\pi^+\pi^- \pi^0 2 \gamma)$
c) M($\gamma \gamma)$ versus
M($\pi^+\pi^-\pi^0)$,
d) M($\pi^+\pi^-\pi^0$) if the $\gamma \gamma$
mass is in the $\eta$ band
(0.48~$\leq$~M($\gamma \gamma$)~$\leq$~0.62~GeV) and
e) M($\gamma \gamma$) if the $\pi^+\pi^-\pi^0$ mass
is in the $\eta$ band
(0.52~$\leq$~M($\pi^+\pi^-\pi^0$)~$\leq$~0.58~GeV).
For the reaction $ pp \rightarrow p_fp_s(\pi^+\pi^- \pi^+ \pi^- \pi^0 \pi^0)$
f) $M(\pi^+\pi^-\pi^0)$ versus
M($\pi^+\pi^-\pi^0)$  and
g) M($\pi^+\pi^-\pi^0$) if the other $\pi^+\pi^-\pi^0$
mass is in the $\eta$ band
(0.52~$\leq$~M($\pi^+\pi^-\pi^0$)~$\leq$~0.58~GeV).
Superimposed as a shaded histogram
are the respective histograms when the central system has a mass greater then
1.3~GeV.
}
\label{fi:1}
\end{figure}
\begin{figure}[h]
\caption{
a) The \etaeta mass spectrum.
Superimposed as a shaded histogram
is an estimation of the background contribution.
b) The background subtracted \etaeta mass spectrum with fit described in
the text.
Superimposed as a shaded histogram
is an estimation of the $f_2(1270)$ and $f_2^\prime(1525)$
contribution.
The c) Real and d) Imaginary parts of the roots ($Z$, see text)
as a function of mass obtained from the PWA of the \etaeta system.
}
\label{fi:2}
\end{figure}
\begin{figure}[h]
\caption{ The $\protect\sqrt{4\pi}t_{LM}$ moments from the data.
Superimposed as a solid histogram
are the resulting
moments calculated from the PWA of the
\etaeta final state.
}
\label{fi:3}
\end{figure}
\begin{figure}[h]
\caption{The physical solution from the PWA of the \etaeta final state.
Superimposed as a shaded histogram
is an estimation of the background contribution.
}
\label{fi:4}
\end{figure}
\begin{figure}[h]
\caption{The background subtracted
a) $S_0^-$ and b) $D_0^-$ waves with fits described in the text.
The parameters used to describe the resonances
are those previously determine from the $\pi\pi$ and
$K \overline K$ channels.
}
\label{fi:5}
\end{figure}
\newpage
\begin{center}
\epsfig{figure=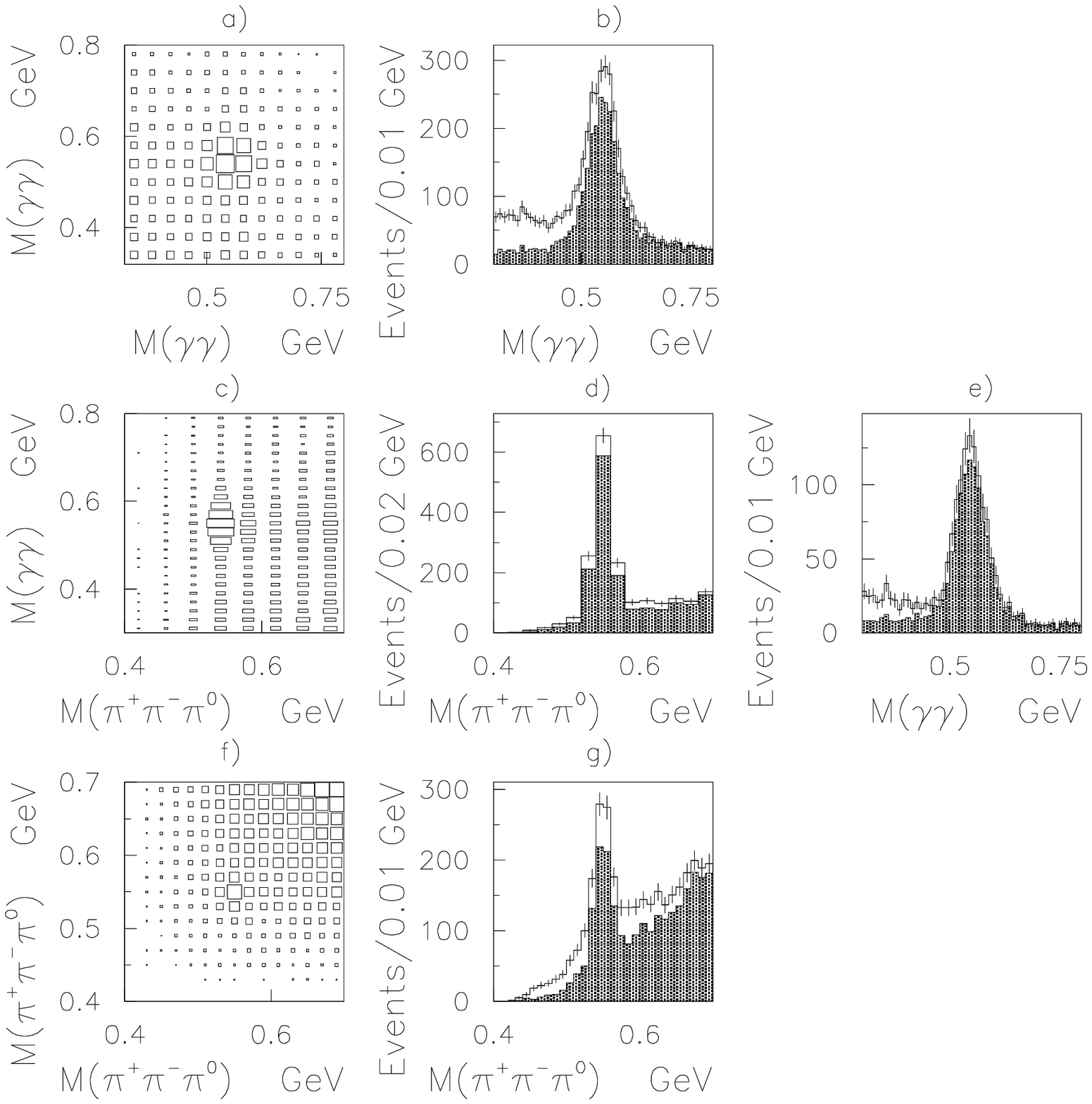,height=22cm,width=17cm}
\end{center}
\begin{center} {Figure 1} \end{center}
\newpage
\begin{center}
\epsfig{figure=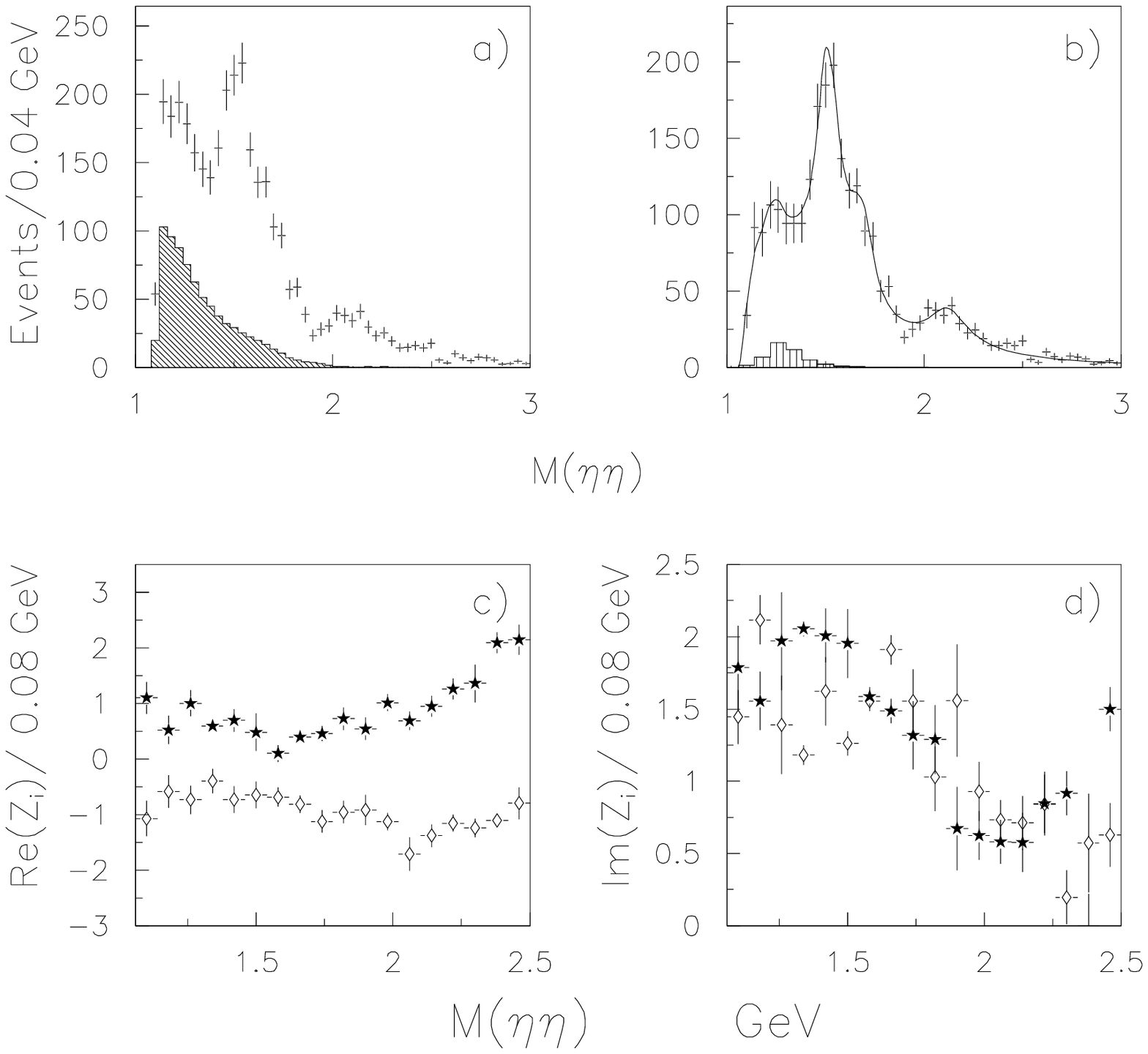,height=22cm,width=17cm}
\end{center}
\begin{center} {Figure 2} \end{center}
\newpage
\begin{center}
\epsfig{figure=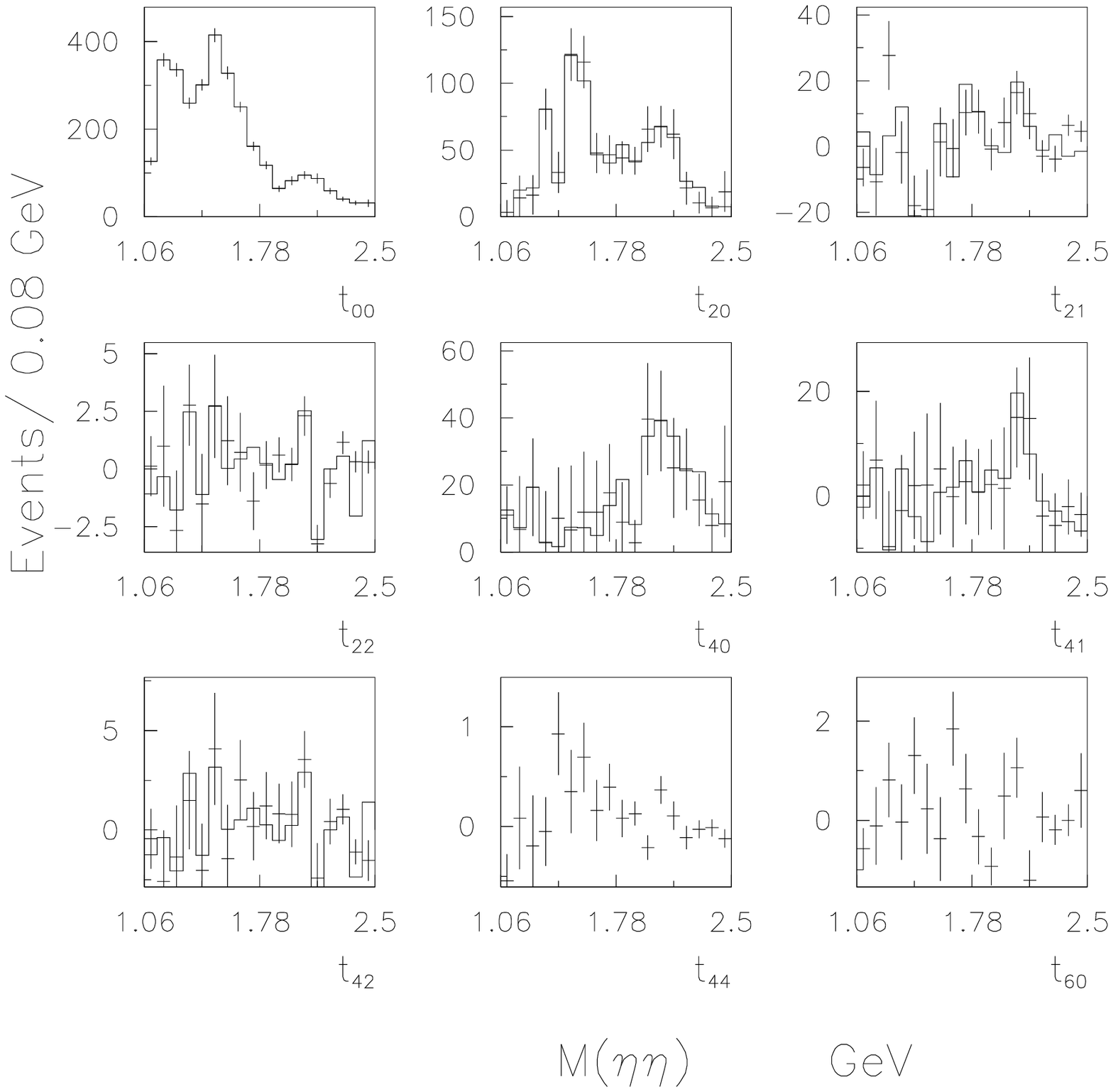,height=22cm,width=17cm}
\end{center}
\begin{center} {Figure 3} \end{center}
\newpage
\begin{center}
\epsfig{figure=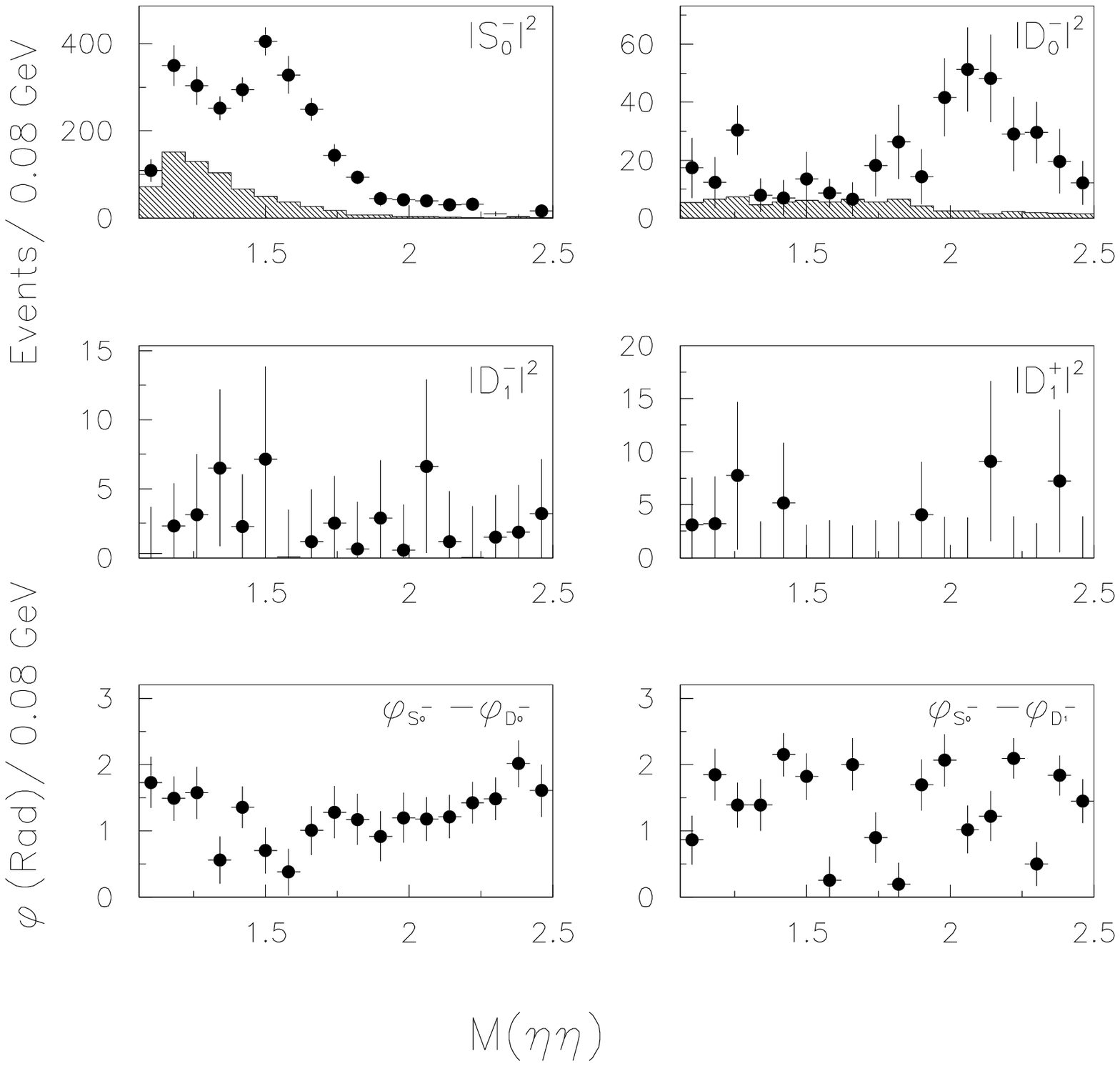,height=22cm,width=17cm}
\end{center}
\begin{center} {Figure 4} \end{center}
\newpage
\begin{center}
\epsfig{figure=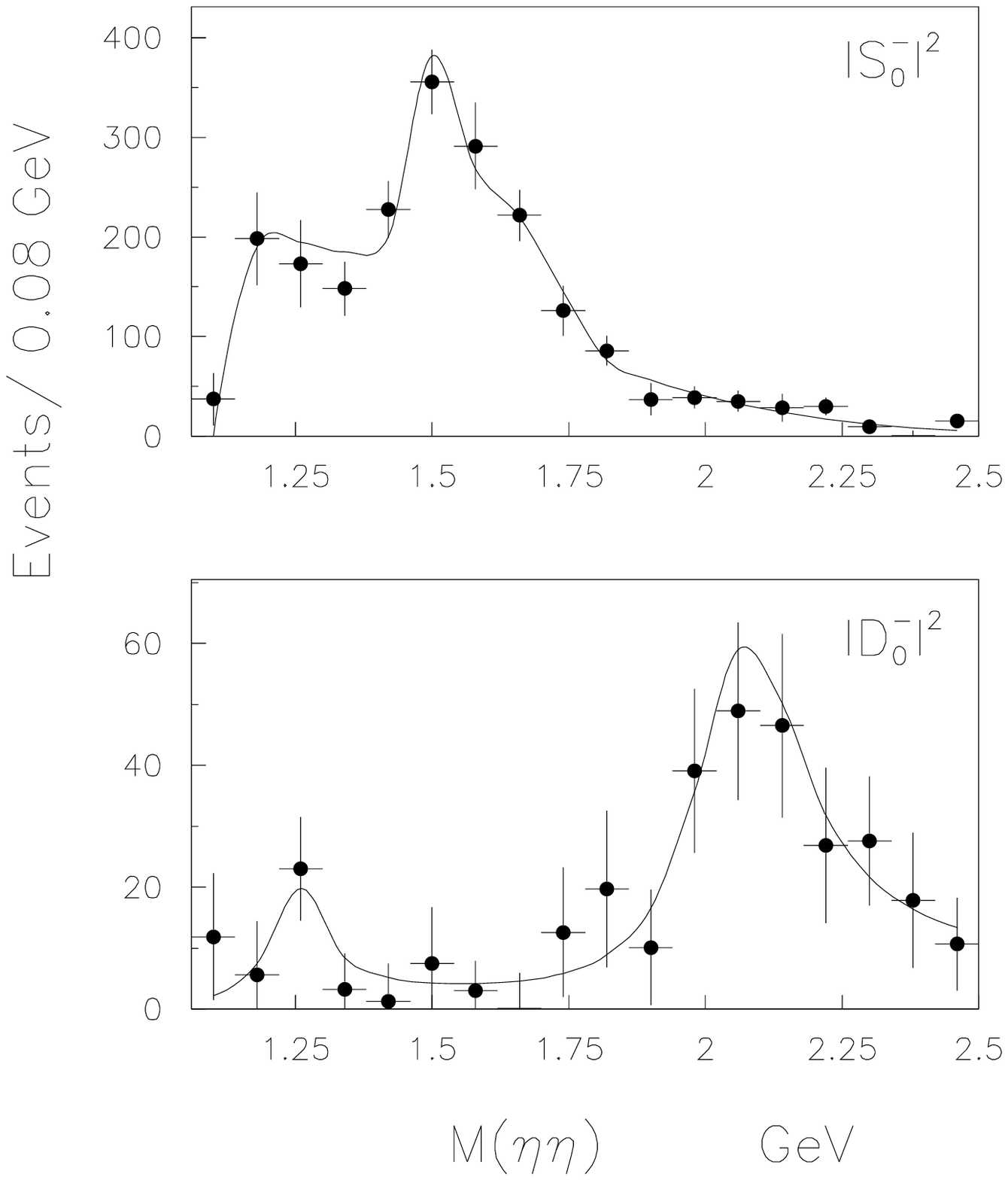,height=22cm,width=17cm}
\end{center}
\begin{center} {Figure 5} \end{center}
\end{document}